\documentclass[runningheads,a4paper]{svproc}

\usepackage{amssymb}
\usepackage{graphicx}

\newcommand{\keywords}[1]{\par\addvspace\baselineskip
\noindent\keywordname\enspace\ignorespaces#1}

\usepackage[pdftex,hyperfootnotes=false,pdfpagelabels]{hyperref}
\usepackage{slashed}
\usepackage[normalem]{ulem}
\usepackage{color}
\usepackage{array}
\usepackage{amsmath}
\usepackage{bbold}

\begin{document}

\mainmatter  % start of an individual contribution

\title{The 3 cavity prototypes of RADES, an axion detector using microwave filters at CAST}

\titlerunning{The 3 cavity prototypes of RADES}

\author{S.~Arguedas Cuendis \inst{1}
\and A.~\'Alvarez Melc\'on \inst{2} \and C.~Cogollos \inst{9} \and A.~D\'iaz-Morcillo \inst{2} \and\\
B.~D\"obrich \inst{1} \and J.D.~Gallego \inst{4} \and B.~Gimeno \inst{5} \and I.G.~Irastorza \inst{3} \and \\
A.J.~Lozano-Guerrero \inst{2} \and C.~Malbrunot \inst{1} \and P.~Navarro \inst{2} \and C.~Pe\~na Garay \inst{6,7} \and J.~Redondo \inst{3,8} \and T.~Vafeiadis \inst{1} \and W.~W\"unsch \inst{1}}
\authorrunning{S. Arguedas Cuendis et al.}

\institute{European Organization for Nuclear Research (CERN), 1211 Geneva 23, Switzerland \\ \and Department of Information and Communication Technologies, Universidad Polit\'ecnica de
Cartagena, Murcia, Spain \and Departamento de F\'isica Te\'orica, Universidad de Zaragoza, 50009, Zaragoza, Spain \and Yebes Observatory, National Centre for Radioastronomical Technologies and Geospace Applications, Guadalajara 19080, Spain \and Department  of  Applied  Physics  and  Electromagnetism-ICMUV,  University  of  Valencia,
Spain \and I2SysBio, CSIC-UVEG, P.O. 22085, Valencia, 46071, Spain \and Laboratorio Subterr\'aneo de Canfranc, Estaci\'on de Canfranc, 22880, Spain \and Max-Planck-Institut f\"ur Physik (Werner-Heisenberg-Institut), 80805 Munchen, Germany \and ICCUB, Universitat de Barcelona, 08028 Barcelona, Spain
}

\toctitle{The 3 cavity prototypes of RADES}
\tocauthor{Authors' Instructions}
\maketitle

\begin{abstract}
The Relic Axion Detector Experimental Setup (RADES) is an axion search project that uses a microwave filter as resonator for Dark Matter conversion. The main focus of this publication is the description of the 3 different cavity prototypes of RADES. The result of the first tests of one of the prototypes is also presented. The filters consist of 5 or 6 stainless steel sub-cavities joined by rectangular irises. The size of the sub-cavities determines the working frequency, the amount of sub-cavities determine the working volume. The first cavity prototype was built in 2017 to work at a frequency of $\sim$ 8.4 GHz and it was placed at the 9 T CAST dipole magnet at CERN. Two more prototypes were designed and built in 2018. The aim of the new designs is to find and test the best cavity geometry in order to scale up in volume and to introduce an effective tuning mechanism.  Our results demonstrate the promising potential of this type of filter to reach QCD axion sensitivity at X-Band frequencies.
\end{abstract}

\section{Introduction}
One of the most important endeavours in physics these days is the search for dark matter. There are a plethora of theories that bring into consideration several possible dark matter candidates. One of these candidates is the axion. Axions, as well as more generic axion-like particles (ALPS), are currently considered one of the most promising fields for new physics beyond the Standard Model (SM). Axions arise in extensions of the SM through the Peccei-Quinn (PQ) mechanism \cite{Peccei:1977hh,Peccei:1977ur}, currently the most compelling solution to the strong-CP problem. \cite{PhysRevLett.40.223,PhysRevLett.40.279}. The exact mass of the axion is not known dependent on its cosmological history. For axion models with PQ transition happening after inflation, some lattice QCD calculations suggest a lower bound to the axion mass of around $\sim$ $m^2_a > 25 \ \mu $eV \cite{Klaer_2017}. 

There is thus theoretical motivation to pursue the search for axions at these masses for which new cavity geometries can be constructed to use as detectors following the conventional axion haloscope technique \cite{PhysRevLett.51.1415}. This technique consists of a high-Q microwave cavity inside a magnetic field to induce the conversion of axions from our galactic dark matter halo into photons. For a cavity whose resonant frequency matches $m_A$, the conversion is enhanced by a factor proportional to the quality factor of the cavity $Q$. So, these type of experiments try to detect axions by the extra power that the converted photon leaves in the cavity. The figure of merit for such an experiment is given by:

\begin{equation}
\label{eq:1}
F \sim g_{A\gamma}^4 m_A^2B^4V^2T_{sys}^{-2}G^4Q \; ,
\end{equation}

where $g_{A\gamma}$ is the axion-photon coupling, $B$ is the magnetic field, $V$ the volume, $T_{sys}$ the system temperature and $G$ is the geometrical form factor of the cavity.

In order to go to higher frequencies (higher masses), one would have to reduce the size of the typical cylindrical cavity used in most haloscope experiments. This will reduce the volume of the detector and at the same time the figure of merit. To tackle this problem, RADES proposed a microwave filter composed of N cavities connected through irises. The working frequency will be determined by the size of the sub-cavity while the volume can be increased with no problems (in principle) by adding more sub-cavities to the filter. Section \ref{section2} will briefly introduce the theoretical model in which the design and construction was based on. Section \ref{section3} will describe the 3 different prototype cavities and their characterization.

\section{Theoretical Model}
\label{section2}
Consider a number of N cavities connected through irises, the cavities have very similar geometries and thus a similar fundamental mode at a common frequency. When excited by a monochromatic axion DM field, the system of coupled equations for the amplitudes of the fundamental mode is given by \cite{Melcon:2018dba}:

\begin{equation}
\label{eq:2}
(\omega^2 \mathbb{1} - \mathbb{M}) \vec{E} = -g_{A\gamma}B_e A_0 \omega^2 \vec{G} \; ,
\end{equation}

where $\vec{E}$ is a vector containing the amplitudes and relative phase of the electric field in each cavity and the matrix $\mathbb{M}$ contains the natural frequencies, damping factors and coupling between the cavities.

It is important to emphasize that the value of the natural frequencies and coupling between the cavities can be arbitrarily chosen by altering the dimensions of the cavities and irises. Equation \ref{eq:2} was solved after the eigenvalue $\omega^2$ was fixed to a desired characteristic frequency. The simplest solution is to take all coupling coefficients to be equal to a fixed value K. The solution showed that all sub-cavities should resonate at the same frequency except for the first and last one. Finally, a filter that maximizes the geometric factor for that frequency was designed. The full derivation and computations mentioned above can be found in \cite{Melcon:2018dba}.

\section{Cavity Prototypes}
\label{section3}
\begin{figure}[]
\centering
\includegraphics[width=0.75\textwidth]{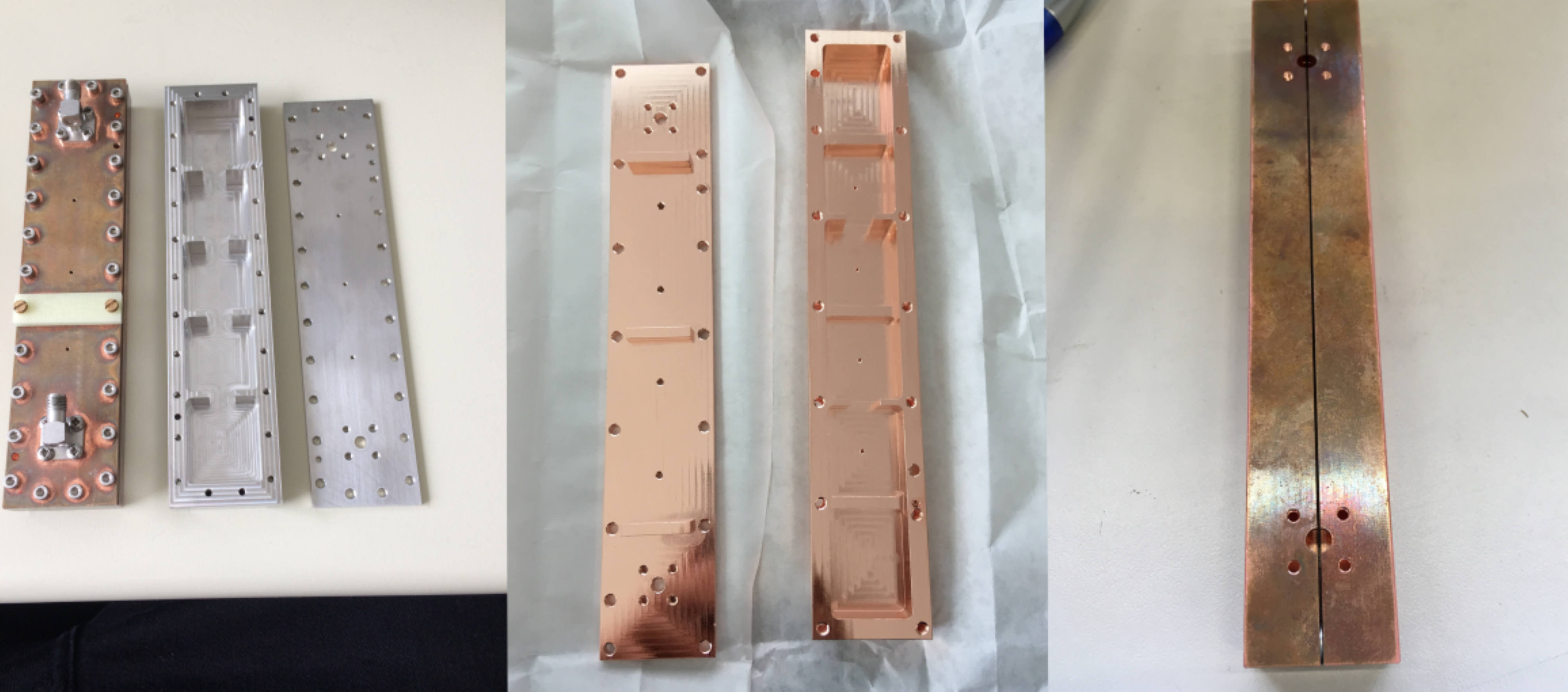}
     \caption{The 3 RADES cavities. Left: Inductive irises prototype. Middle: Alternating irises prototype. Right: Vertical cut prototype.}
     \label{fig:1}  
\end{figure}

RADES has produced so far a total of 3 cavity prototypes. The first one was built in 2017 while the other two were finished in 2018. Fig. \ref{fig:1} shows these three microwave filters , described in the following. The length of the cavities is around 15 cm each.

\subsection{Inductive Irises Prototype}
\begin{figure}[]
\centering
\includegraphics[width=0.6\textwidth]{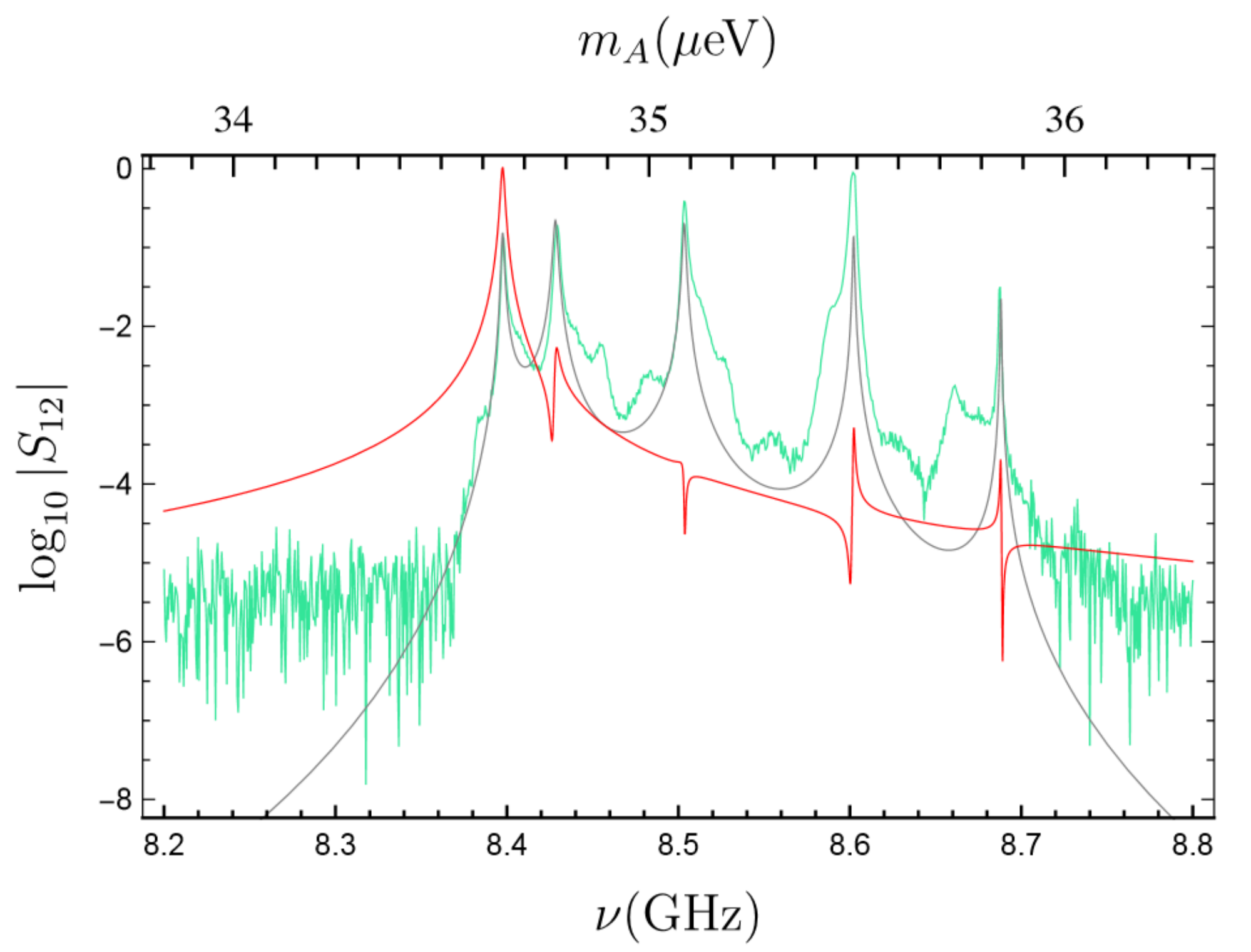}
     \caption{Transmission parameter: measurement (green) at 2K in the CAST magnet and theoretical model (gray). Red is cavity-mode to axion coupling for the 5 modes. Taken from \cite{Melcon:2018dba}.}
     \label{fig:2}  
\end{figure}

The first prototype consists of 5 sub-cavities connected with inductive irises. The cavity was first designed and modeled using CST Microwave Studio using the predictions of the theoretical model to set the dimensions of the sub-cavities. The frequency of operation was fixed to $\omega_1$ = 8.4 GHz and the geometric factor was optimized for this frequency. In 2017, this prototype was successfully installed at the CAST magnet at CERN. Fig. \ref{fig:2} compares the prediction of the theoretical model versus the cryogenic measurement at CAST. Also the strength of the axion coupling is plotted for the different resonances. One can see from this figure that the resonance structure behaved as expected  and the first resonance peak is the one that couples to the axion. Around 350 hours of data has been taken at CAST with this prototype in 2017 and 2018 and the data analysis is ongoing. 

The next step for the collaboration is to increase the volume of the cavity. However, the first simulations with 20 sub-cavities showed that there is a high chance of mode mixing for the first resonance peak due to large amount of resonances. This will reduce drastically the value of the geometric factor. Nevertheless, this can be solved if the working resonance peak is a higher mode. For higher modes, the spacing between adjacent modes is larger and the mode mixing can be avoided. Section \ref{sec.3.2} explains how this can be achieved.    

\subsection{Alternating Irises Prototype}
\label{sec.3.2}
\begin{figure}[]
\centering
\includegraphics[width=0.89\textwidth]{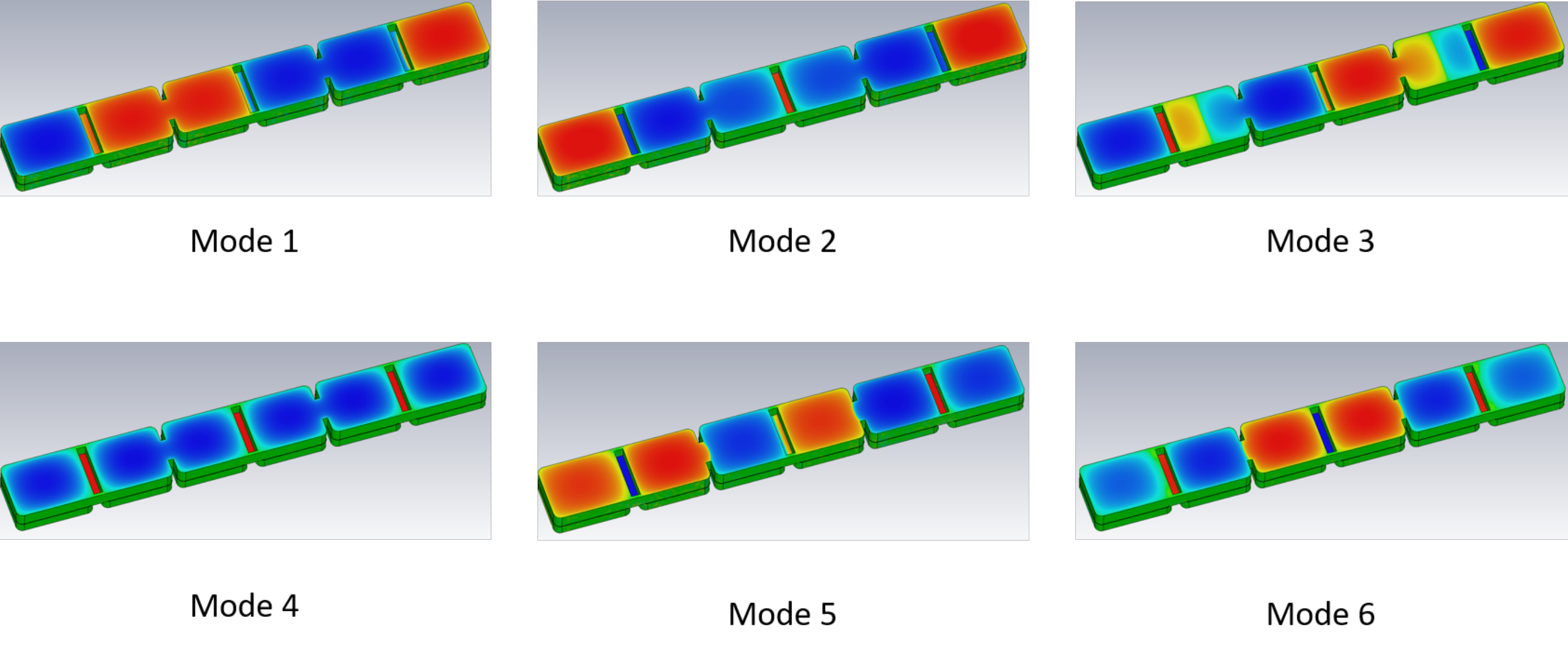}
     \caption{Electric field distribution of the 6 modes in the `alternating irises prototype'.}
     \label{fig:3}  
\end{figure}

The second prototype consists of 6 sub-cavities with inductive and capacitive irises. By introducing capacitive irises, the value of coupling between two cavities can be negative. Solving Equation \ref{eq:2} for alternating irises shows that the axion couples to a higher mode. CST Microwave Studio was used afterwards to design this new prototype. As expected from the theory, the simulations showed that the axion couples to the fourth mode. This can be seen in Fig. \ref{fig:3}: in the fourth mode the electric field in all the sub-cavities is pointing to the same direction. 

This structure was tested at room temperature exhibiting good agreement with expectations derived from the CST model. In the future, tests will be made at 2 K to see the performance of the cavity. It is foreseen to install this cavity (or a 30 sub-cavities structure with alternating irises) at the CAST magnet in summer of 2019 and take data with it.

\subsection{Vertical Cut Prototype}
The final important step to increase the sensitivity range of our detectors would be to tune the resonance frequency. For this reason, a 5 sub-cavities prototype with inductive irises was built. However, this time it was manufactured with a vertical cut, leaving 2 halves that can be separated from each other (see Fig. \ref{fig:1}). By introducing a small gap between the two structures, the width of the structure changes and so does the resonance frequency. 

\begin{figure}[]
\centering
\includegraphics[width=0.75\textwidth]{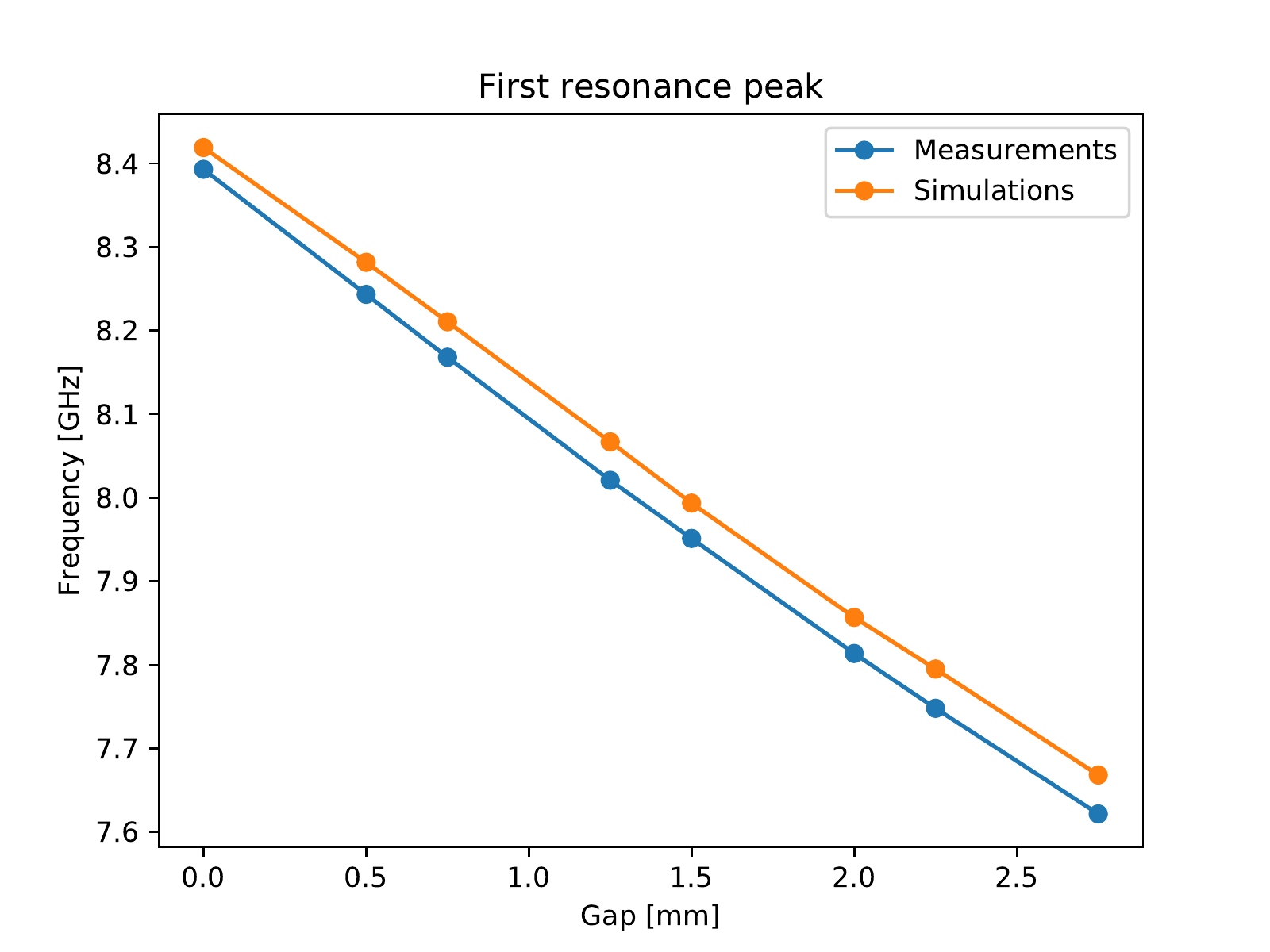}
     \caption{Frequency of the first resonance for different gap sizes. Simulations and measurements.}
     \label{fig:4}  
\end{figure}

Proof of principle tests to tune the cavity were done at CERN, where spacers were manually introduced between the two halves and the resonance frequency was measured using a vector network analyzer. Fig. \ref{fig:4} compares the results of these measurements with the results of the simulations.

First results show a tuning of approximately 800 MHz. At the moment this prototype is sitting at CERN's cryogenic laboratory and is undergoing performances tests at low temperatures (down to 9 K so far).  

\section{Conclusions and Future Plans}
There is convincing theoretical motivation to look for axions with a mass above 25 $\mu$eV. The conventional cylindrical cavity geometry can be challenging when going to higher frequencies. RADES presented a new geometry that allows searches at an axion mass of around 34 $\mu$eV. The theoretical model was developed and then tested using simulations. Afterwards, the first prototype was successfully built, installed and used for data taking at the CAST magnet. New geometries to upgrade the detector were built and are under study at the moment. The first tests of the new structures shows promising results in order to avoid mode mixing and eventually develop tuning for the cavity.

In 2019, the alternating and vertical cut cavities are going to be characterized at 2 K. The construction of an alternating cavity with 30 sub-cavities is planned. This cavity would be installed at the CAST magnet for data taking. A mechanical system to tune the vertical cut cavity will be developed and tested at room temperature. Studies of the possibilities of electrical tuning are also ongoing.     

\section*{Acknowledgment}

This work has been funded by the ERC-STG 802836 (AxScale), the Spanish Agencia Estatal de Investigacion (AEI) and Fondo Europeo de Desarrollo REgional (FEDER) under project FPA-2016-76978.

\bibliographystyle{unsrt}
\renewcommand{\refname}{References}
\bibliography{sample}

\end{document}